\documentclass[prx,aps,twocolumn,superscriptaddress,showpacs,floatfix,amssymb,amsmath]{revtex4-1}
\usepackage{epsfig}
\usepackage{graphics}
\usepackage{subcaption} 
\usepackage{float} 
\usepackage{array}
\usepackage{multirow}

\usepackage{yfonts}

\usepackage{graphicx}
\usepackage{amssymb}
\usepackage{mathrsfs}
\usepackage{bbm}
\usepackage{epsfig}
\usepackage{makecell}

\usepackage[usenames,dvipsnames]{color}

\usepackage{bbm}
\usepackage{color}

\newcommand{\be}{\begin{eqnarray}}
\newcommand{\ee}{\end{eqnarray}}

\begin{document}

\title{
\large %
Rotation by shape change, autonomous molecular motors  \\ and  effective 
timecrystalline dynamics 
      \\
}

\author{Xubiao Peng}
\email{xubiaopeng@bit.edu.cn}
\affiliation{Center for Quantum Technology Research and School of Physics, 
Beijing Institute of Technology, Beijing 100081, P. R. China}

\author{Jin Dai}
\email{daijing491@gmail.com}
\affiliation{Nordita, Stockholm University, Roslagstullsbacken 23, SE-106 91 Stockholm, Sweden}
\author{Antti J. Niemi}
\email{Antti.Niemi@su.se}
\affiliation{Nordita, Stockholm University, Roslagstullsbacken 23, SE-106 91 Stockholm, Sweden}
\affiliation{School of Physics, Beijing Institute of Technology, Haidian District, Beijing 100081, People's Republic of China}
\affiliation{Pacific Quantum Center, Far Eastern Federal University, Vladivostok, Russia}

\begin{abstract}
A deformable body can rotate even with no angular momentum,
simply by changing its shape. A good example is a falling cat,  how it maneuvers 
in air to land on its feet.  Here a first principles  molecular level
example of the phenomenon is presented.  For this the  thermal vibrations of individual atoms
in an isolated cyclopropane molecule are simulated in vacuum and at 
ultralow internal temperature values, and the ensuing molecular motion is followed
stroboscopically. It is observed that in the limit of
long  stroboscopic time steps the vibrations combine into an apparent 
uniform rotation of the  entire molecule  even in the absence of angular 
momentum. This large time scale rotational motion is then modeled  in an effective theory
approach, in terms of timecrystalline Hamiltonian dynamics.
The phenomenon is a temperature sensitive  measurable. As such it has potential applications that range
from models of autonomous molecular motors to  development of 
molecular level detector, sensor and control 
technologies.
\end{abstract}
\maketitle

%
%
%
%

\section{Introduction}

Geometric mechanics  states that  in the case of a deformable body, the vibrational and rotational motions are not
separable  \cite{Guichardet-1984,Shapere-1989a,Shapere-1989b,Littlejohn-1997,Marsden-1997,Katz-2019,Dai-2020b}.
Even with no angular momentum, local vibrations in parts of the body can become self-organized into an emergent,  global
rotation of the entire body.  This effect is already used  for a variety of control purposes, for example the attitude of 
satellites is often controlled by periodic motions of parts of a satellite such as spinning rotors \cite{Marsden-1997}.

Here  the first molecular level realization of such an effective rotational motion is presented. The example  
describes how a molecule employs
thermal  (or quantum) fluctuations of its individual atoms to rotate akin a molecular motor,  even when the molecule does no carry any
angular  momentum.  For this kind of an effective rotational motion to emerge,  the molecule needs to have at least three 
movable components \cite{Guichardet-1984}.  A good example that is described here, 
is a single cyclopropane (C$_3$H$_6$) molecule. It  is
a ring molecule that is made of 
three methylenes (CH$_2$) at the corners of an equilateral triangle. As such, cyclopropane is the simplest example of an
organic  ring molecule. It is being studied widely, as a building block of organic synthesis.  In  medicine 
cyclopropane is also used extensively as an anesthetics \cite{Schneider-2014,Kulinkovich-2015}. 

Other examples of simple triangular molecules that can be analyzed similarly
include aziridine, oxirane and 1,2 dimethyl\-cyclo\-pro\-pane  \cite{Reddy-2018}.

\section{Theory}

To understand the geometric provenance  of deformation induced rotation \cite{Guichardet-1984},  
consider the (time) 
$t$-evolution of three equal mass pointlike objects,  at the corners $\mathbf r_i$  ($i=1,2,3$) of a triangle.
There are no external forces, thus the center of mass  is stationary at all times
\begin{equation}
\mathbf r_1 (t) +  \mathbf r_2 (t) + \mathbf r_3 (t) = 0
\label{r123}
\end{equation}
The total angular momentum  also vanishes
\begin{equation}
\mathbf L \ = \ \mathbf r_1 \wedge \dot {\mathbf r}_1 + \mathbf r_2 
\wedge \dot {\mathbf r}_2 + \mathbf r_3 \wedge \dot {\mathbf r}_3 \ = \ 0
\label{Lz}
\end{equation}
Nevertheless the triangle can rotate, by shape changes. To describe this rotation, 
the triangle is oriented  to always lie on the $z=0$ plane;  
two triangles have the same shape when they only differ  by a rigid rotation on the plane, around the $z$-axis. 
The shape is described
by assigning internal shape coordinates 
$\mathbf s_i(t)$ to the three corners.  These coordinates 
describe all possible triangular shapes in an unambiguous fashion when one demands that 
$\mathbf s_1(t)$ always coincides with the positive $x$-axis so that $s_{1x}(t) >0$ and $s_{1y}(t)=0$, 
and one sets $s_{2y}(t)  >0$. The components of $\mathbf s_3(t)$ are determined as in  (\ref{r123}),
\[
\mathbf s_3(t) =  - \mathbf s_1(t) - \mathbf s_2(t)
\] 

Now let the $\mathbf s_i(t)$ evolve arbitrarily, but in a 
$T$-periodic manner 
\[
\mathbf s_i(t+T) = \mathbf s_i(t)
\] 
The triangular shape then  traces a closed loop 
$\Gamma$ in the space of all possible triangular 
shapes.   At each time the shape coordinates  $\mathbf s_i(t)$ relate to 
the space coordinates  $\mathbf r_i(t) $ by a spatial rotation on the $z=0$ plane,
\begin{equation}
\left( \begin{array}{c} r_{ix}(t) \\ r_{iy}(t) \end{array} \right) = 
\left( \begin{array}{cc}  \cos \theta(t)  &  - \sin \theta(t) \\  \sin \theta(t)  & \  \cos \theta(t) \end{array} \right) \left( \begin{array}{c} s_{ix}(t) \\ s_{iy}(t) \end{array} \right) 
\label{theta}
\end{equation}
Initially $\theta(0)=0$ so that whenever $\theta(\mathrm T)\not= 0$ the triangle has rotated. 
This rotation angle is evaluated  
by substituting (\ref{theta}) into (\ref{Lz}):  
\begin{equation}
\theta(\mathrm T)   \ = \ \int\limits_0^{\mathrm T} \! dt \, \, \frac{
\sum\limits_{i=1}^{3}  \left\{ s_{iy} \dot s_{ix}  - s_{ix} \dot s_{iy}\right\}  
}
{ 
\sum\limits_{i=1}^3 \mathbf s^2_i
}
 \ \equiv \ \int_\Gamma d { \mathbf l } \cdot {\mathbf A}  
\label{dotheta}
\end{equation}
Whenever (\ref{dotheta}) does not vanish the triangle returns to its initial shape but with a net rotation by $\theta(\mathrm T)\not=0$ 
around the axis through its  center of mass. 
Since  there is no angular momentum the rotation is emergent,  it is entirely due to shape changes. 

In (\ref{dotheta}) the angle has also been expressed as a line integral of a connection  
one-form $\mathbf A$
\cite{Guichardet-1984,Shapere-1989a,Shapere-1989b,Littlejohn-1997,Marsden-1997,Katz-2019,Dai-2020b}.  
To identify it, start from Jacobi coordinates 
\[
\begin{array}{ccc} 
\mathbf s_1 &=& \frac{1}{\sqrt{2}} \mathbf u - \frac{1}{\sqrt{6}}  \mathbf v
\\
\mathbf s_2 &=& \sqrt{ \frac{2}{3}}  \mathbf v
\\ 
\mathbf s_3 &=& - \frac{1}{\sqrt{2}}\mathbf u - \frac{1}{\sqrt{6}}  \mathbf v
\end{array}
\label{jacobi}
\]
With
\[
\left( \begin{matrix} u_1 \\ u_2 \end{matrix} \right) = 
r \cos\frac{\vartheta}{2}  
\left( \begin{matrix} \cos\phi _1  \\  \sin\phi_1  \end{matrix} \right) \ \ \ \& \ \ \ 
\left( \begin{matrix} v_1 \\ v_2 \end{matrix} \right) = 
r \sin\frac{\vartheta}{2}  
\left( \begin{matrix} \cos\phi _2  \\  \sin\phi_2  \end{matrix} \right)
\]
and with 
\[
\phi_{\pm} = \phi_1 \pm \phi_2
\]
the connection  one-form $\mathbf A$ becomes
\begin{equation}
\mathbf A  =  - \frac{1}{2} \cos \vartheta  d\phi_-  -  \frac{1}{2} d\phi_+   =   
\frac{1}{2}  \frac{ \ x d y - y dx } {r (r+z) }   -  \frac{1}{2}  d (\phi_+  + \phi_-)
\label{deltatheta}
\end{equation}
Remarkably $\mathbf A$ is akin the connection one-form of a Dirac monopole at the origin $r=0$ of $\mathbb R^3$ with the 
Dirac  string placed along the negative $z$-axis.
Thus (\ref{dotheta}) 
evaluates the flux of the monopole 
through a surface with  boundary $\Gamma$ in the space of triangular shapes. Such a  flux  is generically nonvanishing so that
$\theta(\mathrm T) \not=0$ generically. 

Notably, the Dirac string concurs with a shape where two corners overlap, and at the location of the monopole all three corners of 
the triangle coincide.

\section{Simulation results}

In the case of  a cyclopropane molecule the angle of rotation (\ref{dotheta}) can be 
evaluated numerically, using state-of-the-art all-atom classical molecular dynamics. This is a
methodology that is designed for a realistic, accurate description of a molecular system  
at a semiclassical level. 
The simulation results  presented here are obtained using 
the GROMACS package \cite{gromacs}  with CHARMM36m  force 
field  \cite{charmm36m}  in combination with  {\it CGenFF}  \cite{Vanommeslaeghe-2010,Vanommeslaeghe-2012} for 
parameter generation (simulation details are described in Appendix). 
The boundary conditions are free and  both Coulomb and Van der Waals interactions extend
over all atom pairs in the molecule, with no cut-off approximation.  
The  initial cyclopropane structure is taken from the PubChem web-site \cite{PubChem}. It  
has  $\mathrm D_{3h}$ molecular symmetry, with atomic 
coordinates that are specified with $10^{-14} \ m$ precision.  This approximates the  minimum
of the CHARMM36m potential energy with a precision that even exceeds its intended accuracy.

A set of  50 ``random" initial configurations for the production runs is first constructed, by energy drift  \cite{gromacs}
using single precision floating point format.  Each initial configuration has a different 
internal molecular temperature value  $T$ below $1.0$K that is determined by the equipartition theorem.
The pool of initial configurations is designed to minimize the mechanical CHARMM36m free energy 
at the given internal temperature value $T$, with single precision floating point format.   
In particular, the angular momentum of each initial configuration vanishes with comparable numerical 
accuracy.

%

In the production runs double precision floating point format is used in combination with  
$\Delta \tau =1.0f$s time steps, which is quite standard in high precision all-atom simulations: This is sufficient to stabilize 
the internal molecular temperature so that $T$ remains essentially
constant during each of the production runs, with vanishingly small energy drift.  
Each of the  50 production run trajectories simulate 
10$\mu$s {\it in silico}  and a single trajectory takes about 90 hours of wall clock time
with the available processors.

\begin{figure}
		\includegraphics[width=0.45\textwidth]{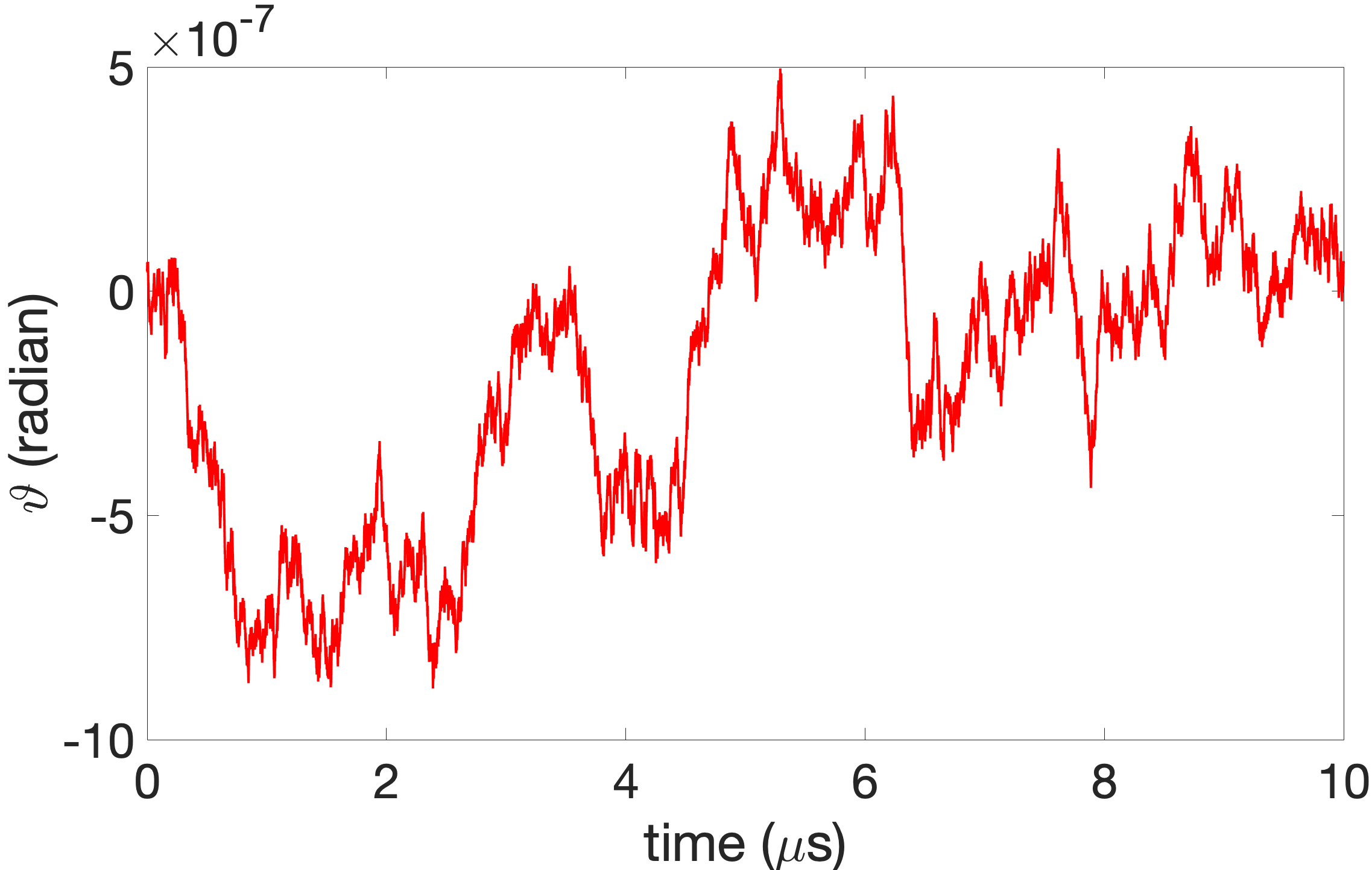}
 		\centering
 		\caption
		{Generic time evolution of the angle $\vartheta(n)$ in equation (\ref{vartheta}) in the production runs: There is no observable 
		net rotation due to angular momentum.  In this example the 
		internal temperature is  $T=0.065$K and the same trajectory is  
		analyzed in Figures \ref{fig-3}.}
  \hfill
  \label{fig-1}
\end{figure}

It  is verified that the angular momentum remains vanishingly small during all of the 
production runs.  For this,  the instantaneous positions and velocities of all the 
individual C and H atoms are recorded at every femtosecond time step $n$ of all production run
trajectories. The instantaneous values $L_z(n)$  of angular momentum component that is normal to the plane of 
the three carbon atoms is evaluated for the entire molecule, together with the corresponding instantaneous 
moment of inertia values $ I(n)$.  This yields the instantaneous 
angular velocity values 
\[
\omega(n) = L_z(n) / I(n)
\] 
These instantaneous values  $\omega(n)$ are summed up, to compute the
accumulated ``rigid body'' rotation angle, that is due to angular momentum: 
\begin{equation}
\vartheta(n) = \omega(n) \Delta \tau + \vartheta(n-1) \ = \  \sum\limits_{i=1}^{n} \frac{L_z(i)}{I(i)} \Delta \tau  
\label{vartheta}
\end{equation}
In full compliance with the condition (\ref{Lz}), in all the production run simulations
the values of (\ref{vartheta}) always remain less than $\sim 10^{-6}$ radians, 
for all time steps $n$ and Figure \ref{fig-1} shows a typical example: In the production simulations there
is no observable rotation of the cyclopropane molecule, due to angular momentum.    Accordingly any
systematic rotational motion that exceeds $\sim 10^{-6}$ radians during a production run must be emergent, 
and  entirely due to shape deformations. 
%
%
%
%
%
%
%
%
%


To observe  such a systematic rotational motion the orientation of the cyclopropane 
structure is monitored stroboscopically,  using 
time steps that are large in comparison to the frequency of a typical covalent bond oscillation.    
It is found that the cyclopropane  molecule always rotates, and the rotation is always around the center of mass 
axis that is normal to the plane of the three carbon atoms. Moreover, the angular velocity of the effective rotational motion
is {\it very} sensitive to the value $T$ of internal molecular temperature. 

\begin{figure}
		\includegraphics[width=0.45\textwidth]{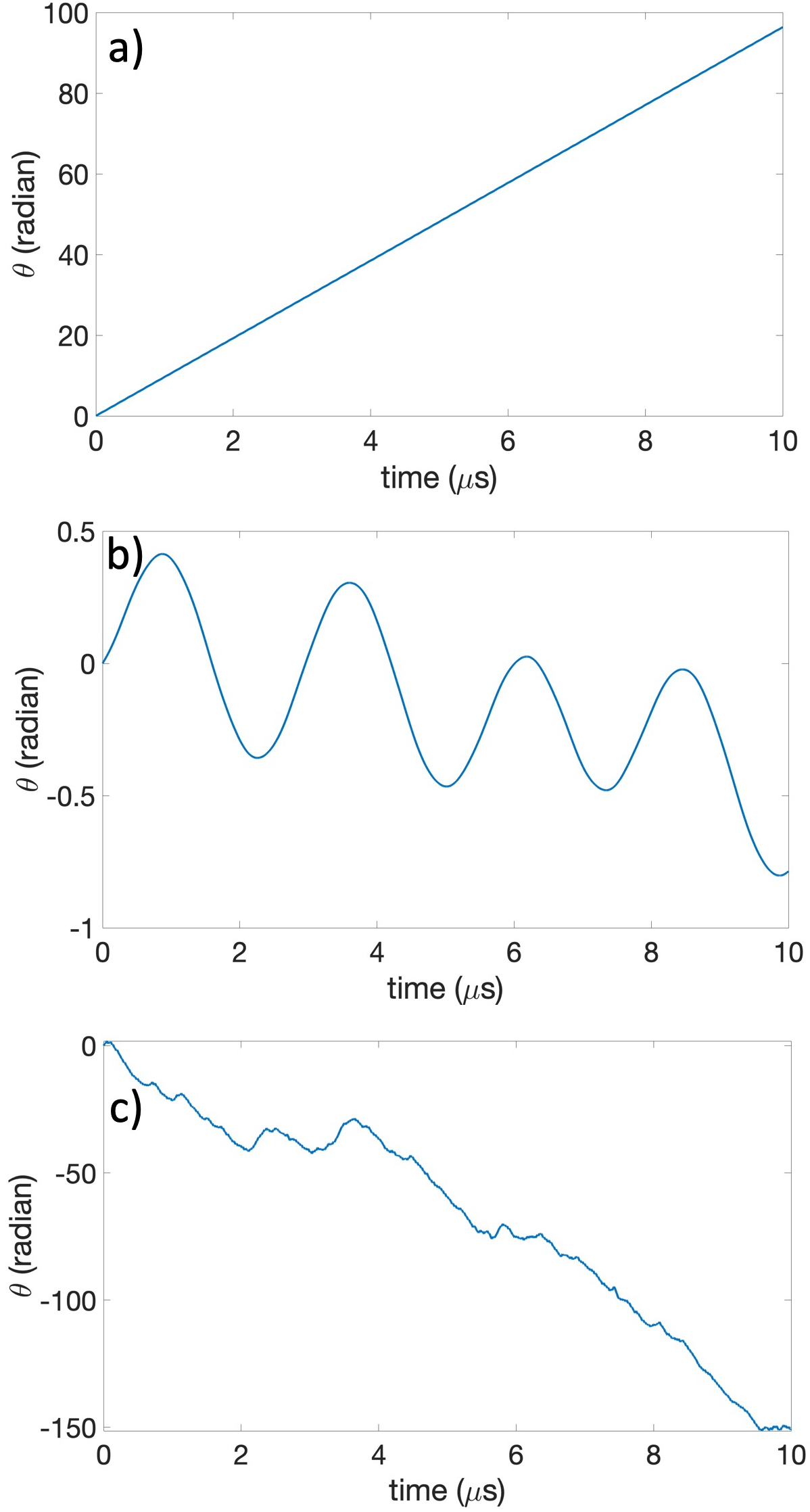}
 		\centering
 		\vskip -0.cm 
 		\caption{Evolution of rotation angle  (\ref{dotheta}) in cyclopropane, 
		at different internal $T$-values  and 
		stroboscopic time step  $\Delta t = 10^{-10}$s. 
		In Figure a) $T=0.066 $K, in Figure b) $T=0.0087$K, in Figure c) $T=0.94 $K.
		See Supplementary material. }
  \hfill
  \label{fig-2}
\end{figure}

In Figures  \ref{fig-2} three different, generic  examples of production run trajectories are presented; 
movies of these trajectories are available as  Supplementary Material. 
In each trajectory, the molecule is observed with 100$p$s  stroboscopic time steps. 

\begin{figure}
		\includegraphics[width=0.45\textwidth]{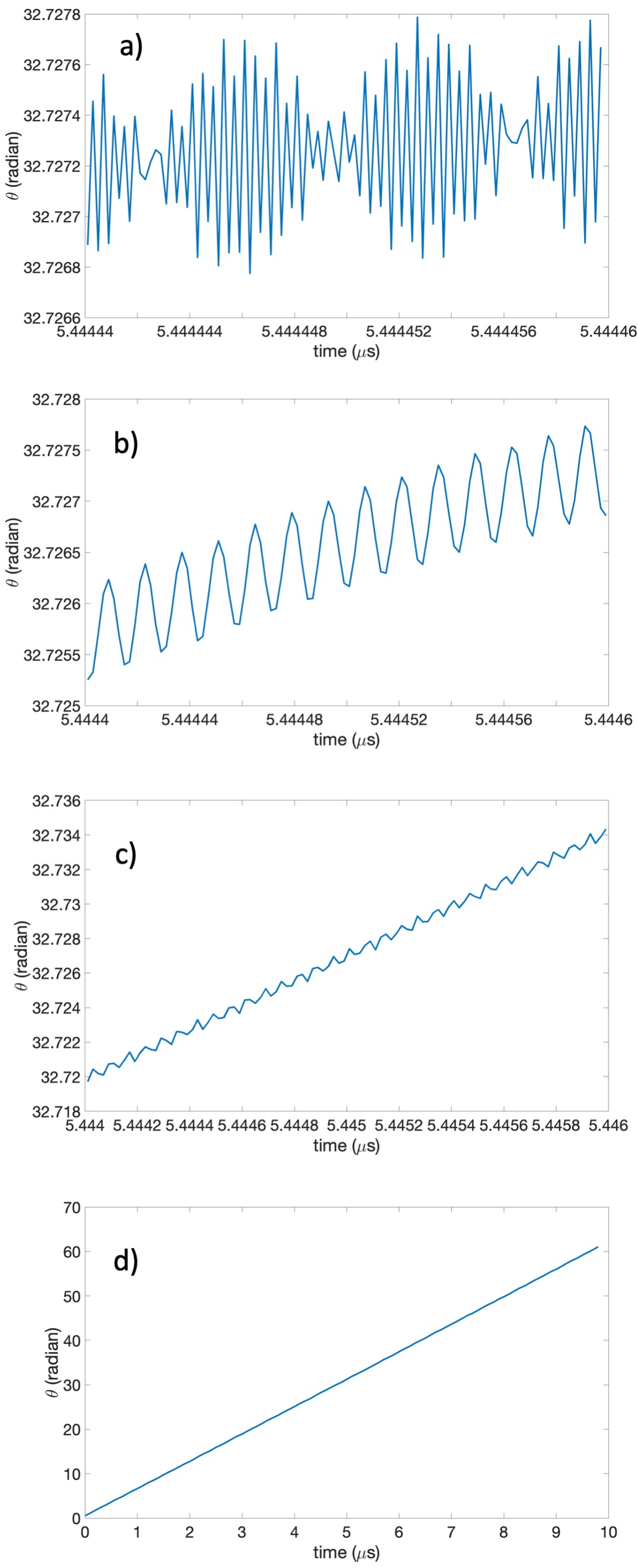}
 		\centering
 		\vskip -0.cm 
 		\caption{Time evolution of rotation angle $\theta(t)$ along a $T=0.065$K trajectory  
		sampled different stroboscopic time steps $\Delta t $.  In Figure a) 
		$\Delta t = 10^{-13}$ s, in Figure b) $\Delta t = 10^{-12}$s, in Figure c) $\Delta t = 10^{-11}$s and
		in Figure d) $\Delta t =  10^{-7}$s.
		}
  \hfill
  \label{fig-3}
\end{figure}

In Figure  \ref{fig-2} a) the internal molecular temperature has the value $T=0.066 $K.   
The molecule rotates  in a uniform fashion clockwise {\it i.e.} with increasing $\theta(t)$ 
at a  constant rate of $\sim$10 rad/$\mu$s around the  axis that is normal to the plane of the carbon.
In Figure  \ref{fig-2} b) $T= 0.0087$K. Now the rotational motion is ratcheting, with a drift towards 
decreasing values of $\theta(t)$ at around $\sim$0.07 rad/$\mu$s. Such a combination of ratcheting and drifting
motion is quite commonly observed, at low $T$-values and with those relatively short stroboscopic time 
steps that are attainable in the simulations.    
The drift implies that with larger stroboscopic time steps,  in excess of 10$\mu$s which is the length of a production 
trajectory, the rotation becomes increasingly uniform; this will be exemplified in Figures \ref{fig-3}.
Finally, in  Figure  \ref{fig-2} c) $T=0.94 $K which is close to the upper bound of $T$-values accessible in the present 
simulations,  while maintaining a good numerical stability.  Now the molecule rotates in the  counterclockwise direction 
at a  rate around $\sim$15 rad/$\mu$s during the  10$\mu$s simulation. The rotation remains slightly 
erratic at the stroboscopic time scale. Presumably, an increase in the stroboscopic 
time scale in combination with a longer MD trajectory produces a uniform  rotation also in this case.
 
Unfortunately, with the available computers it is not feasible to try and produce 
trajectories that are much longer  than 10$\mu$s, with a high level of numerical stability. 
Such trajectories would be needed,  to analyze in detail higher $T$-values.  The present simulation results propose 
that there is a critical value $T_c$ probably around  a few Kelvin,  above which uniform stroboscopic  
rotation can no longer be observed. This could be a consequence of simulation
artifacts that are due to issues such as energy drift  \cite{gromacs}.

%
%
%
%
%

In Figures \ref{fig-3} the trajectory shown in Figure \ref{fig-1} is analyzed in more detail;  the internal temperature value 
$T=0.065 $K in this Figure is very close to the $T$-value of  the
trajectory in Figure \ref{fig-2} a). 
%
%
%
%
%
%
In Figure  \ref{fig-3} a) $\theta(t)$  is sampled with very short  stroboscopic time steps $\Delta t = 200 $$f$s.  
Oscillatory rotational motion is observed,  with an amplitude that also oscillates but at much lower frequency.
In Figures  \ref{fig-3} b) and  \ref{fig-3} c) the  time steps are increased 
to $\Delta t = 2.0 $$p$s and  $\Delta t = 20 $$p$s, respectively.
Now an increasingly regular ratcheting motion is observed, with apparent clockwise rotation, and  with a decreasing relative amplitude; 
see also Figure \ref{fig-2} b). 
Finally, in Figure  \ref{fig-3} d) where  $\Delta t = 100 n$s  the cyclopropane structure apparently rotates and in a uniform manner,
akin the  rotational motion shown in Figure \ref{fig-2} a).   
Despite a very small difference in $T$-values, 
the angular velocity is now clearly slower. Indeed, a peak of angular velocity that is centered at $T\approx 0.066 $K is very
narrow. Thus the rotational motion is a very sensitive measurable. 

\section{Effective theory analysis}

The Figures \ref{fig-3} demonstrate how an emergent separation of scales leads to a dynamical self-organization as  the 
stroboscopic time steps increase. In the final Figure \ref{fig-3} d) the cyclopropane rotates like an equilateral triangle, in a uniform fashion at
constant angular velocity,  around the symmetry axis that is normal to the plane of the three methylenes. 
This emergent uniform rotation has all the characteristics of energy conserving, Hamiltonian dynamics.  As such, it can be described by an effective 
Hamiltonian theory in terms of a reduced set of variables, appropriate to govern the molecule in the limit of large stroboscopic time steps.  
It is apparent  that this  effective theory must coincide with the 
timecrystalline Hamiltonian model analyzed in \cite{Dai-2019,Alekseev-2020,Dai-2020b}.  This is an effective theory
that describes the time evolution of three point-like
interaction centers $\mathbf x_1, \mathbf x_2 $ and $ \mathbf x_3 $ in terms of the three link vectors 
\[
\mathbf n_i = \mathbf x_{i+1} - \mathbf x_i
\] 
with
$\mathbf x_4 = \mathbf  x_1$ that are subject to the Lie-Poisson brackets 
\[
\{ n^a_i , n^b_j \} = \pm \delta_{ij} \epsilon^{abc} n^c_i
\]
where the two signs are related by parity. Hamilton's equation is
\begin{equation}
\frac{ \partial \mathbf n_i } { \partial t } \ = \ \{ \mathbf n_i , H(\mathbf n) \} \ = \ \mp  \mathbf n_i \times  \frac{ \partial H}{\partial \mathbf n_i}  
\label{Hameq}
\end{equation}
Since 
\[
\{ n^a_i , \mathbf n_j \cdot \mathbf n_j \} = 0 
\]
for all $i,j$ the bracket preserves the length of $\mathbf n_i$ independently of the Hamiltonian;
for clarity all $| \mathbf x_{i+1} - \mathbf x_i | $ are then chosen equal so that the structure is an equilateral triangle. 
Indeed, the bracket is
designed to generate any kind of dynamics of the vertices $\mathbf x_i$ except for stretching and shrinking of the 
links.   
With  Hamiltonian
\[
H \ =  \ \mathbf n_2 \cdot (\mathbf n_1 \times \mathbf n_{3} )
\]
Hamilton's equation (\ref{Hameq}) describes uniform rotation of the  triangle, with constant 
angular velocity around the symmetry axis that is normal to the triangle. This is exactly what is encountered in the case of a 
cyclopropane, at large stroboscopic time steps as shown in Figures \ref{fig-2} a) and \ref{fig-3} d). 
It follows from  \cite{Dai-2019,Alekseev-2020,Dai-2020b} that cyclopropane is a realization of the time crystal phenomenon \cite{Wilczek-2012,Shapere-2012,else-2016,Yao-2017,zhang-2017,choi-2017,Sacha-2018,Nayak-2019,Shapere-2019,Yao-2020}.

\vskip 0.2cm
The two rotation directions around the equilateral symmetry axis of the cyclopropane are related by
parity. In the present simulations the direction becomes selected randomly, essentially by floating point round-up errors, while in
the effective Hamiltonian theory the direction changes with the sign in  the Lie-Poisson bracket. 
In the case of actual cyclopropane, the molecule is quantum mechanical and it displays anomalous energetic and  magnetic behaviors that are 
suggestive of a ring current \cite{Dewar-1984,Fowler-2007,Sundholm-2016}.  More generally, the  ability to 
sustain a diamagnetic (diaptropic) 
ring current in the presence of an external magnetic field is often considered as a  defining characteristics of aromatic ring 
molecules \cite{Schleyer-1996,Gomes-2001}.  Ring current is a quantum manifestation 
of an effective electron current flow that appears in a semiclassical treatment of the molecule,  
and all-atom molecular dynamics captures this current flow as an effective counter-rotational motion of the nuclear backbone. 
In the absence of a magnetic field, the two directions of electron current flow specify two 
degenerate quantum mechanical states. The ground state is a positive superposition of the two, and it is 
separated by a small energy gap from the negative superposition. An applied external magnetic field selects an energetically  
preferred direction for the current to flow. This is the essence of spontaneous  symmetry breakdown.

\section*{Summary and outlook}

In summary, the article describes results from 
state-of-art all-atom molecular dynamics simulations that model the ground state properties of a single
cyclopropane molecule in vacuum, at ultralow  internal molecular temperature values.  Even when there is no angular momentum, 
an emergent ratcheting rotational motion 
is commonly observed. This effective rotational motion is due to a collective self-organization of  individual atom thermal vibrations,  it
takes place at 
characteristic time scales that are large in comparison to the frequency scale of covalent bond oscillations. When 
the molecule is followed with sufficiently long stroboscopic time steps the emergent rotation appears uniform, in a manner that
can be reproduced  by an effective theory timecrystalline Hamiltonian dynamics. The angular velocity of the 
rotation is found to be very sensitive to the internal molecular temperature. In the case of a single cyclopropane molecule studied here,
the ability to produce uniform rotational motion without angular momentum appears to be limited to 
ultralow internal molecular temperature values.  But in the case of a larger molecular system the pertinent temperature scale for rotation
should increase with increase in the molecular size. 

The emergent timecrystalline 
rotation could  be commonplace in the case of aromatic ring molecules. As a manifestation of aromatic ring current
it could even serve as their defining characteristic. More generally, vibration induced rotational motion could conceivably fuel molecular motors. 
Its apparent high temperature sensitivity 
proposes that the phenomenon could be employed for a variety of detection, sensor and control 
purposes,  in situations where extremely high  precision is desirable.   The phenomenon could even govern the {\it in vivo} working of
biological macromolecules.

\section*{Acknowledgements}

AJN thanks Frank Wilczek for numerous clarifying discussions on all aspects of the work.
XP is supported by Beijing Institute of Technology Research 
Fund Program for Young Scholars. DJ and AJN are supported  
by the Carl Trygger Foundation Grant CTS 18:276, by the Swedish Research Council under Contract No. 2018-04411, and by COST Action CA17139.
AJN is also partially supported by Grant No. 0657-2020-0015 of the Ministry of Science and Higher Education of Russia.

\section*{Appendix}

The all-atom molecular simulations are performed with the GROMACS package \cite{gromacs}. The  
force field is CHARMM36m  \cite{charmm36m} and parameters are generated using   {\it CGenFF} \cite{Vanommeslaeghe-2010,Vanommeslaeghe-2012}. 
A single cyclopropane molecule is simulated in vacuum under NVE conditions: There is no ambient temperature or pressure 
coupling in the simulations. 
The  boundary conditions are free. Both Coulomb and Van der Waals interactions extend over all atom 
pairs in the molecule with no cut-off approximation. The initial cyclopropane structure is taken from the PubChem web-site \cite{PubChem} where it is specified with $10^{-14} \, m$ precision in both the C and H atomic coordinates.

The simulation process starts with refined search of minimum energy configuration 
of the potential energy contribution in CHARMM36m free energy. For this the initial PubChem configuration 
is subjected to dissipation with 100  consecutive  GROMACS simulations.  
Each of the dissipation simulations use  the final coordinates of the atoms 
obtained in the previous simulation as its initial configuration, but with all velocities 
of all atoms set to zero.  Each of the 100 trajectories has a  length of  1.0 picoseconds ($p$s), and the time 
step is 0.01 femtoseconds ($f$s). Double precision floating point format is used 
in all these simulations.
At the end of the 100$^{th}$ simulation the atoms are static, 
the structure is fully $\mathrm D_{3h}$ symmetric and in particular
there is no kinetic energy left, with double precision accuracy.
The configuration is then even further refined, using  the L-BFGS 
optimization algorithm \cite{Byrd-1995} until the maximal force becomes 
less than $10^{-12}$ kJ/mol/nm.  
Finally,  it is confirmed that the structure
is indeed a locally stable potential energy minimum,  by performing a 
10$\mu$s double precision simulation with 1.0$f$s time step 
and vanishing initial velocities. During this entire simulation the total energy is 
conserved,  there is no movement of the atoms observed.
This confirms that the configuration is a static $\mathrm D_{3h}$ symmetric
local minimum energy of the potential energy, with double precision accuracy.

The final configuration specifies all the atomic positions with better than 
$\Delta x = 10^{-24} \, m$ double precision accuracy. Similarly, all 
the atomic velocities vanish with better than $\Delta v = 10^{-11} \, m$s accuracy.  
Thus, both in the case of hydrogen and
carbon atoms $\Delta x \Delta p << \hbar$ and as such the structure 
is specified much more precisely than should be reasonable, by  quantum mechanical considerations.  
To alleviate this,  and in order to reach sufficiently long molecular dynamics trajectories, 
the investigation proceeds with single  precision floating point format; even with  single precision 
accuracy the atomic positions are determined with a precision that is better than {\it e.g.}
the proton radius. A  single precision simulation is initiated from the double precision
minimum energy configuration, using a 1.0$f$s time step. 
It is observed that during the first few time steps, the single precision potential energy slightly
decreases from its initial value. At the same time, all the covalent bond lengths start to 
oscillate. It is concluded that this instability of the initial double precision minimum energy 
configuration is  due to  accumulation of round-up errors. It is 
caused  by an incompatibility of the  single precision and double precision floating point 
formats in combination with CHARMM36m parameter values,
in the present case. After the first few initial steps during which the potential energy 
decreases,  the total energy starts increasing  but at a very slow rate:  The single precision 
simulation is subject to a very slow energy drift, a phenomenon that  
is quite common in all-atom MD simulations.  See GROMACS 
manual for detailed description \cite{gromacs}.

When the covalent bonds oscillate and the atoms are  in motion, equipartition theorem is used
to assign the entire molecule an internal temperature. A GROMACS routine is used to estimate the temperature.
This energy drift of the single precision simulation is employed,  
to prepare the cyclopropane molecule at a desired value of internal temperature $T$. For this  
the single precision simulation is extended, until a desired internal temperature value has been reached.  
From that  point on, the simulations are  continue with double precision,  for the production runs that extend a
full 10$\mu$s of {\it in silico} time, with 1.0$f$s time step.  
During these simulations only insignificant energy drift is observed, the internal temperature is very stable. 
For example, in the case of the $T=0.065K$ simulation described in Figures 3 of main text the energy difference between the final
state, and the initial double precision ground state, is  $\Delta E =  1.244 \times 10^{-4}\, $eV; with $\Delta t = 10^{-15}$ time steps this gives
$\Delta E \Delta t << \hbar$.

\vskip 0.2cm
We note that for a proper analysis a 
full all-atom quantum molecular dynamics should be
performed. Unfortunately,  such a simulation is out of reach for  presently available computers.

%
%
%
%
%
%
%
%
%

\vskip 1.0cm

\vskip 0.4cm

\section*{Supplementary material}

\noindent
The supplementary material consists of three movies, 
corresponding to the three panels of Figure 2 in main text. Each 33 second movie describes the 
entire 10 microsecond simulation trajectory.  

\vskip 0.5cm

$\bullet ~$ Movie 1 is trajectory with $T = 0.066$K \\
\vskip 0.2cm
$\bullet ~$ Movie 2 is trajectory with $T =0.0087$K \\
\vskip 0.2cm
$\bullet ~$ Movie 3 is trajectory with $T = 0.94$K

\end{document}